# Design of Field Programmable Gate Array (FPGA) Based Emulators for Motor Control Applications

Ahmed Ben Achballah, Slim Ben Othman and Slim Ben Saoud
Advanced Systems Laboratory – Polytechnic School of Tunisia
National Institute of Applied Sciences and Technology – Department of Electrical Engineering
B.P. 676, 1080 Tunis Cedex, Tunisia

**Abstract: Problem Statement:** Field Programmable Gate Array (FPGA) circuits play a significant role in major recent embedded process control designs. However, exploiting these platforms requires deep hardware conception skills and remains an important time consuming stage in a design flow. High Level Synthesis technique avoids this bottleneck and increases design productivity as witnessed by industry specialists. **Approach:** This study proposes to apply this technique for the conception and implementation of a Real Time Direct Current Machine (RTDCM) emulator for an embedded control application. **Results:** Several FPGA-based configuration scenarios are studied. A series of tests including design and timing-precision analysis were conducted to discuss and validate the obtained hardware architectures. **Conclusion/Recommendations:** The proposed methodology has accelerated the design time besides it has provided an extra time to refine the hardware core of the DCM emulator. The high level synthesis technique can be applied to the control field especially to test with low cost and short delays newest algorithms and motor models.

## INTRODUCTION

The high integration scale of FPGAs and their high-speed processing time with the reconfigurability option make this type of circuits an attractive solution for many types of applications (Rodriguez-Andina *et al*., 2007). FPGA based systems and simulators are found in various domains like defense (Gonzalez *et al*., 2008), medical (Monmasson and Cirstea 2007), renewable energy (Ouhrouche 2009), physics and the control of industrial process (Ben Salem *et al*., 2010; Idkhajine *et al*., 2008; Naouar *et al*., 2007). However, implementing complex algorithms in FPGA-based systems can be a laborious work. This study is still realized by circuit-vendor specific tools in many cases and requires deep design skills, so it remains the most time consuming operation in a design flow (Gupta *et al*., 2004; Paiz *et al*., 2008).

Modern design tools give the possibility to designers to overcome reconfigurable circuit limits and to shorten the product availability in the markets. Some of these tools are called high level compilers, frameworks and also synthesizers, derived from the word "High Level Synthesis". This technique consists of the translation of an algorithm from a high level language like C to an equivalent hardware language like VHDL or Verilog that represents a circuit description (R. Coussy and Morawiec, 2008; Pellerin and Thibault 2005; Philippe C. and A. Morawiec, 2008). The resulting hardware descriptions can be implemented directly into circuits like FPGAs. Hardware engineers may not modify them and so economize in the design process time (Martin and Smith 2009). Such time gain can be used in the test and the on-chip verification steps. A recent study in the industrial field demonstrates that HLS technique is necessary to increase productivity and diminish the gap between the increasing integration of chips and the number of designers needed to work on them ( Philippe C. and A. Morawiec, 2008). The same study shows that hardware engineers who tried the HLS technique wouldn't leave it because of its performance and practical obtained results. By this way, the industry of electrical process control, which is yet beneficiary from FPGAs advanced platforms, is now benefiting from advancements of EDA tools and techniques including HLS. The application of such advantages (circuitry and tools) in the process control field is generally concentrated on implementing more efficient complex algorithms and on testing them in real conditions with an association with a motor model. This technique is often called Hardware in the Loop simulation (HiL) (Dufour *et al*., 2007). The proposed FPGA-based simulators are various and depend on parameters like 1) The degree of algorithms" complexity (controller and motor model) 2) Computing accuracy 3) Timing constraints.

**Corresponding Author:** Ahmed Ben Achballah, LSA Laboratory–EPT/INSAT, B.P. 676, 1080 Tunis Cedex, Tunisia

These parameters oblige designers to follow different design flows (Martin and Smith 2009) and to find alternative solutions especially against timing constraints like Real-Time Operating System (RTOS), Multi-Processors System On Chip (MPSoC) or the hardware implementation of several parts of the electrical process (e.g., controller unit). Still also, some design methodologies which are dedicated to the hardware design field, are more and more adopted and applied to the process control one like hardware-software CoDesign using languages like SystemC or SpecC (Salewski and Kowalewski 2008).

In this study, we explore HLS technique because, as we know, it was not already applied to the embedded control domain. In fact, we investigate the efficiency of using the benefits of this technique to automatically generate a hardware module of the RT DCM emulator circuit. The purposes behind this case study are 1) to evaluate this technique with a basic electrical motor model design 2) to extract the advantages and the disadvantages of applying HLS techniques to the control domain.

**Related works:** FPGA based systems approaches on the industrial control field were often based on similar tools and environments. Among them, Xilinx System Generator (XSG) from Xilinx Inc., DSP Builder from Altera and SymplifyDSP from Synopsys were very solicited from the research community.

In (Monmasson and Cirstea 2007), the XSG tool was used to implement an FPGA-based controller for AC drives and where two case studies were presented. Reference (Paiz *et al*., 2008) introduced an enhanced simulation board dedicated to the rapid prototyping of digital controllers and also used one of the cited tools above to generate hardware descriptions from high level descriptions.

The literature also contains alternative approaches, Opal-RT team proposed a Real-Time simulation platform RT-XSG (including model's libraries) to perform Hardware-in-the-Loop (HiL) simulation of electrical drives but it still depends on XSG tool to complete the synthesis and the implementation of the targeted FPGA circuit (Dufour *et al*., 2008). HiL testing phase is essential in the validation process of control units and motor drives. In addition, prior works are focused on two major axes 1) FPGA implementation of complex control algorithms for performance purposes 2) Validation of electrical controllers and/or motors at earlier stages of production for cost reasons (Martin and Smith 2009). These approaches don"t consider the validation and the diagnosis of the electrical process after the production stage.

To resolve this problem, the emulation approach can be a solution for testing control algorithms. This concept is assured by the addition of a new validation stage between simulation and experimentation. After the validation of the control unit, commands can be directly applied to the real motor avoiding its destruction which could be expensive and factor of delayed delivery of the product (Braham *et al*., 1997). Once the emulation performed successfully, the designed emulator can be used for diagnostic applications. The development of such emulators is essentially faced with the execution time problems since its main function is to reproduce real systems behavioral that are highly dynamic.

Such approaches were already developed by (Ben Othman *et al*., 2008; Ben Salem *et al*., 2008). The last two studies have been applied to the Real-Time emulation of an embedded controller for a DC Machine but they have been conducted using a pure software solution or a mixed one (software and hardware). Despite the fact that the emulator execution time in last two approaches was competitive, it doesn't allow the recuperation of instantaneous values (below 1 µs computing steps) and so, not enough closer to a real motor functioning. This limitation can induce more penalties due to the evolution of control algorithms and the complexity of some electric machine models.

For this reason, our approach is to design a hardware module which operates as a co-processor to the on-chip processor. This will economize software delays such as 1) The sequential execution of software instructions inside the processor 2) Interruptions and context switching latencies and as a result, accelerate the emulator computing time to meet timing constraints. Because the hardware computing is faster than the software one, the purpose of the emulator hardware implementation is to obtain less than 1 µs computing steps (called hor). With this computing time, the emulator can intimately reproduce the functioning of a real motor. Also, recent digital motor controllers can have a sampling rate below 10 µs, so it was inevitable for us to maintain a competitive computing time for the simulated Direct Current Machine (DCM) (Dufour *et al*., 2007).

Meanwhile, the design stage has to be quick, flexible and reproducible in case of eventual modifications on the emulator model. HLS technique is a good candidate to ensure the last cited conditions. In what remains, we will detail our approach by the application of this technique to the emulation concept of a DCM process, followed by the obtained results and the discussion.

**Emulation concept:** Emulating electrical systems are to reproduce their functionalities with the most accurate model in a virtual manner. The goals behind such approach is allowing designers to validate control units and to diagnose them after in terms of precision and efficiency (Ben Achballah *et al.*,, 2010; Ben Saoud *et al.*, 1996). These two stages are described in the following two paragraphs.

**Validation Stage:** In an emulation approach, the validation stage came as the first step to elaborate. In fact, the controller is associated to a motor emulator that represents the real motor and controller commands are validated against it. This maneuver is executed before the association of the controller with the real motor. The connection between these two elements has to be similar to the real one and assure a realistic data exchange. The controller sends command signals to the emulator that, in his turn, reproduces information about the motor state as they are issued from sensors. When the control device is validated it can be switched for application to the physical motor as (Fig. 1).

**Diagnosis Stage:** Simultaneously, the same commands applied to the real process are also applied to the emulator. The output signals of the emulator constantly compare to the real motor ones. Using the received data from both terminals, we can analyze the behavior of the real system and detect irregular functioning (Fig. 2). Note that if the emulator"s output is speed; it has to be limited because the emulator is not looped back.

**Realization constraints:** The realization of real time emulators is closely related to the adopted design methodology and the used technology. In fact, these two factors have a direct impact on the performances of the designed emulator. The first one, if not enough specified, can induce to an inadequate architecture to the technology that will encapsulate the emulator later.
Also, it can conduct to a complicated design flow that can increase the conception time. The second factor, the used technology, depends itself on many other parameters. Among them we can cite hardware platforms (microcontrollers, DSPs, FPGAs, ASICs) or CAD tools (design, simulation, on-chip verification.

The implementation process is faced to a one major constraint which is the execution time of the emulator's algorithm. This factor will allow the evaluation of the emulator's performances including its capability to reproduce the real process. In the following, we will introduce the emulated process considered in this case study.

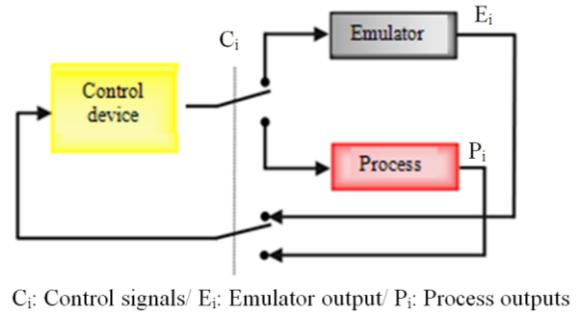

$C_i$: Control signals/ $E_i$: Emulator output/ $P_i$: Process outputs

Fig. 1: Structure of the validation application

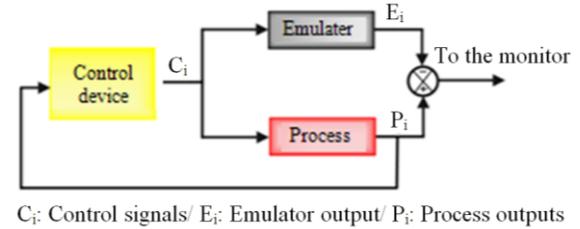

$C_i$: Control signals/ $E_i$: Emulator output/ $P_i$: Process outputs

Fig. 2: Structure of the diagnostic application

## MATERIALS AND METHODS

**DCM Process:** we have chosen to study a direct current motor case because it is a simple electrical machine model. The purpose behind this study is to test the HLS approach in the control domain field (Ben Achballah *et al.*,, 2010).

The model we propose is by two elements which are the control unit and a DC motor emulator. In addition, a chopper is utilized to aliment the emulator with Vh voltage. Two parameters are furnished by the emulator to the controller which are the motor current ($I_m$) and speed ($\Omega_m$). Another parameter is also considered by the control unit which is reference speed $\Omega_{Ref}$ entered by users to supply the adequate duty cycle alpha to the chopper module. The system is demonstrated in Fig. 3 while its parameters are resumed in Table 1. To compute the system state, we use basic mathematical models for a DC electric motor; the chopper (Eq. 1), the current (Eq. 2) and the rotation speed (Eq. 3). The equation parameters are recapitalized in Table 2.

$$Vh = (2.alpha - 1).Vin \qquad (1)$$

$$\frac{dim}{dt} = \frac{1}{Lm}(Vh - Em - Rm.Im) \qquad (2)$$

$$\frac{d\Omega m}{dt} = \frac{1}{J}(Cem - Cr) \qquad (3)$$

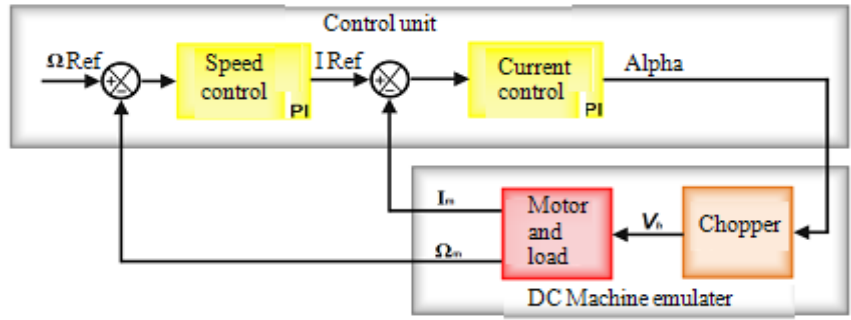

Fig. 3: General diagram of the DCM process

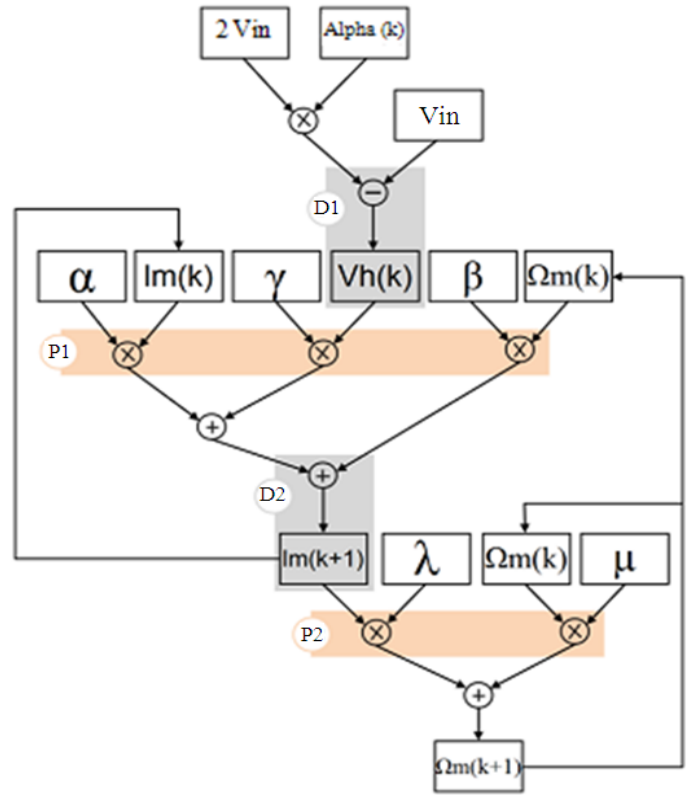

Fig. 4: DFG execution of the emulator

Where:

$Em = Km \cdot \Omega m$
$Cem = Km \cdot Im$
$Cr = K1 \cdot \Omega M^2 \cdot Sign(\Omega m) + K2 \cdot \Omega m + K3 \cdot Sign(\Omega m)$

**DCM emulator algorithm:** The equivalent algorithm of the studied process is performed using mathematical models (Eq. 1, 2, 3) with second order Runge-Kutta sampling method. It is given by (Eq. 4, 5, 6):

$$Vh(k) = (2 * alpha(k) - 1) * Vin \quad (4)$$

$$Im(k+1) = a * Im(k) + \beta * \Omega m(k) + \gamma * Vh(k) \quad (5)$$

$$\Omega m(k+1) = \lambda * Im(k) + \mu * \Omega m(k) + \nu * sign[\Omega m(k)] \quad (6)$$

Where $\alpha$, $\beta$, $\gamma$, $\lambda$, $\mu$ and $\nu$ are calculated from system parameters and the computing step hor.

Table 1: Control unit parameters

| Parameter | Nomenclature |
|---|---|
| Vin | Supply voltage |
| Vh | Chopper's output voltage |
| Em | Back-electromotive force |
| Im | Machine current |
| Ωm | Machine rotation speed |
| Cem | Electromagnetic torque |
| Cr | Resistant torque |
| Lm | Inductance |
| Rm | Resistance |
| J | Inertia |
| Km | Electromagnetic torque coefficient |
| K1, K2, K3 | Resistant torque coefficient |

Table 2: Equations parameters nomenclature

| Parameter | Value |
|---|---|
| Current controller gains kp, kpi | 1.1737, -1.0150 |
| Speed controller gains kp, kpi | 0.142, -0.1111 |
| Iref Limits | ± 13 A |
| Current sampling time | 300 μs |
| Speed sampling time | 20 ms |
| PWM frequency | 16 kHz Design 2 (page 12) |
| Dead time | not used |

Table 3: DCM Algorithm parameters

| Parameter | Value |
|---|---|
| α | 0.9995 |
| β | -9.1977e-005 |
| γ | 4.9987e-004 |
| λ | 1.4603e-004 |
| μ | 1 |
| ν | 0 |
| Vin | 60 Volts |
| hor | 350 ns (page 7) |

The equivalent data flow graph (DFG) of the emulator's algorithm is shown in Fig.4 (for $v = 0$).

Although the dependence between the algorithm equations variables (D1-D2), parallelism can be extracted from the execution flow (P1 and P2) to reduce computing time. This concerns the multiplications in Eq. 5 and 6 where the computation can be assured by independent hardware multipliers for each one.

The result is a faster execution time but this will increase the area consumption in the targeted FPGA circuit especially when the computing is realized with floating point arithmetic. For this case study, we will focus on generating parallelized architectures to gain in speed because it is our primary concern. However, we will evaluate the quality of the obtained circuits in term of area consumption.

The algorithm parameters used after for the simulation and more lately in the hardware implementation tests are provided in Table 3.

**HLS Approach an overview:** To face the increasing integration capacity of chips and the customer's insatiable demand of complex applications, development of Electronic Design Automation (EDA) tools and methodologies have to find innovative solutions (Pellerin and Thibault 2005). HLS technique is one of among available solutions which were kept by both academicians and industrials.

As a proof, we can invoke some recent experiments conducted by three leading industrial companies and explaining that HLS tools have to be considered in the future for cost and productivity reasons Philippe C. and A. Morawiec, 2008). However, this success is the result of many critiques that followed HLS tools since their arrival on the market (Martin and Smith 2009).

In fact, the efficiency of such environments in terms of area consumption, the control of hardware generation flow and the quality of the final design was enormously discussed. By this way, several studies were conducted to evaluate different HLS environments against diverse criteria. As an example we cite the BDTI program.

Nowadays, HLS tools are more and more mature to be considered by industrial society. We just cite a few of them, CoDeveloper from ImpulseC, DK Suite, CatapultC... The field of application varies depending on the purpose behind the use of such technique. One of the scientific fields which benefitted from HLS tools is the signal processing domain, where parallelism is massively extracted and then allows the computing acceleration, sometimes hundreds of times.

Unfortunately, control algorithms and machine models have the characteristics to be variable-dependent and so, could be difficult to automatically extract parallelism from them. In our case study, we will investigate the use of CoDeveloper as HLS environment to generate a hardware module of a DCM emulator. In the following paragraphs, we will introduce this tool and expose obtained preliminary simulation results.

**CoDeveloper high level synthesis tool and flow:** CoDeveloper is an HLS tool developed by Impulse Accelerated Technology. It is based on ImpulseC language which is its input language based itself on Stream-C environment developed in the Los Alamos labs.

To utilize CoDeveloper, developers have to follow its programming approach which is based on Communication Sequential Processes (CSP). In other terms, a set of C functions that represent software or hardware modules connected with data channels (Fig. 5). Data channels are composed of data streams, signals or registers.

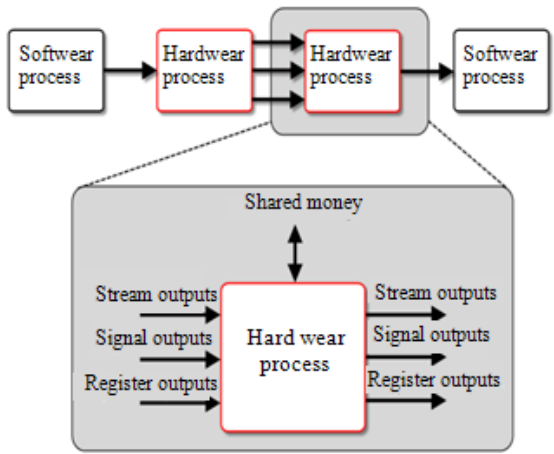

Fig. 5: CSP programming model of ImpulseC

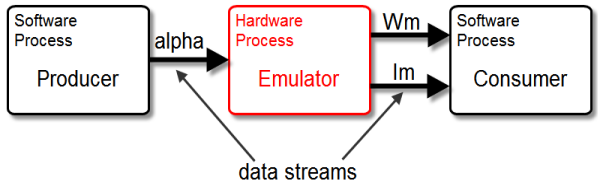

Fig. 6: Processes partition of the DCM emulator

Software processes are used as generators to furnish simulation scenarios or as consumers to collect resulting data from the simulation. They can also play the role of drivers if the hardware module is associated with a processor. In the other side, hardware modules are converted to a hardware description in an xHDL language. CoDeveloper includes three internal tools which are:

- Application Monitor simulates the design.
- CoValidator generates xHDL testbench files.
- StageMaster explores the design for a step by step verification.

**CSP programming model of the DCM emulator:** To generate VHDL description from C codes, designers have to follow the CSP programming model and to convert their C application to ImpulseC syntax. In this case, the emulator application was divided into 3 processes as explained in Fig. 6. The linking between them is assured by 32 bits width data streams. This choice is based on two parameters 1) All data are in floating point format 2) The emulator module will be implemented as a co-processor to the MB processor with FSL connections (Ben Achballah *et al.*,, 2010).

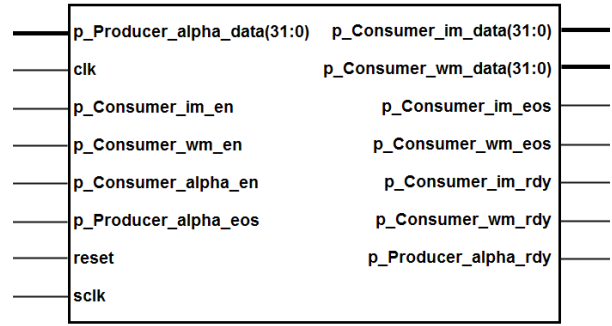

Fig. 7: Hardware module of the DCM emulator

Table 4: CoDeveloper report of the emulator hardware process

| Hardware resources | | Timing analysis | |
|---|---|---|---|
| Operators | Used | Total stages | Max. Delay (In clock cycles) |
| Floating-point Adders/Subs tractors (32 bits) | 4 | 55 | 32 |
| Floating-point Multipliers (32 bits) | 5 | | |
| Estimated DSPs (18×18 Multipliers) | 20 | | |

The software processes (Producer and Consumer) are necessary to simulate the hardware process (Emulator) with the Application Monitor tool. The hardware generation is realized according to this configuration 1) Xilinx MicroBlaze FSL (VHDL) 2) Double-precision types and operators. The obtained circuit after the HLS of the hardware process from ImpulseC code is shown in Fig. 7.

CoDeveloper generates also an estimation report of hardware resources consumption and timing analysis. This report is summarized in Table 4.

**Simulation results:** We proceed to the simulation of the DCM emulator circuit. This simulation concerns:

- The accuracy of computed values (Im and Ωm represented respectively by "im-data" and "wm-data") and which is verified by Application Monitor tool.
- Timing analysis to measure the execution time of the DCM emulator circuit and which is verified by ModelSim tool.

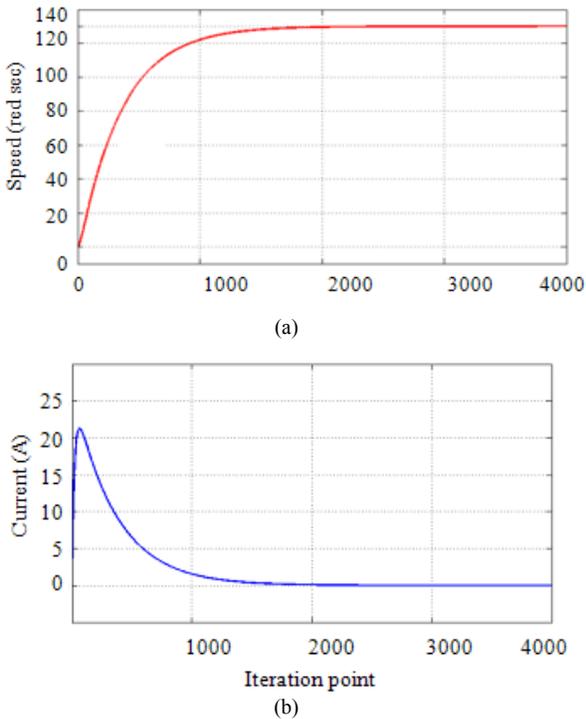

(a)

(b)

Fig. 8: PC-based Speed (a) and Current (b) responses of the DCM emulator

**Computing results:** the alpha level used in this simulation is 0.5- 0.7 at zero. Theoretically, it induces a stationary rotation speed of 130.1308 rad/s. In this stage of simulation, the current and the speed curves are expressed in terms of iteration points and all calculations are realized in floating point format. The emulator responses are plotted in Fig. 8.

**Execution time results:** the execution time of the DCM emulator was measured using "ready flags" among other useful signals which are automatically generated by CoDeveloper (Fig. 7 for port details). We just name two of them:

- xxx_rdy: indicates that input or output data xxx_data are ready in the corresponding streams
- xxx_en: can be used to control the data transferring in the streams

In this measure, we obtained 350 ns computing steps for the DCM emulator. By applying a system clock of 100 MHz, this time corresponds to which was provided in the estimation report in Table 4 (Approximately 32 clock cycles). This time will also allow the high fidelity reproducing of a real motor functioning. The simulation measurement results are illustrated in Fig. 9.

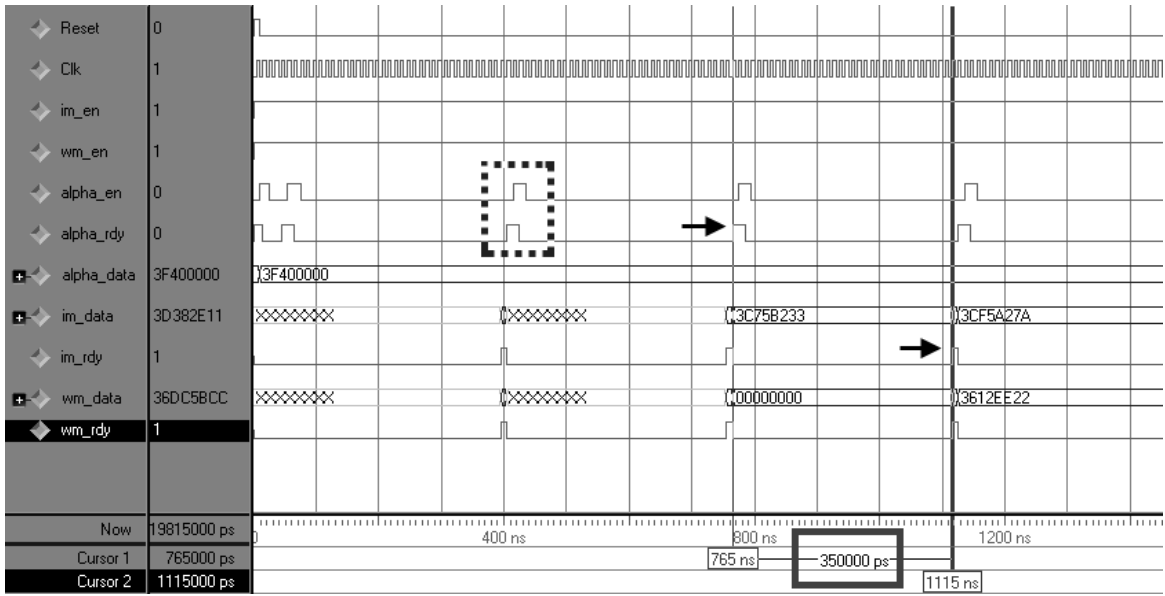

Fig. 9: Simulation measures of the computing time for the DCM emulator circuit

## RESULTS

We present experimental results and analysis obtained from the implementation of the RT DCM emulator in the FPGA circuit. To evaluate the system two tests are considered: The first is on-chip verifications of execution time and computing results. For the second test, the DCM controller is added to the system to complete the HiL platform and where the speed and current responses are studied. Finally, an extra validation test is also realized and where a controller failure is inserted into the emulating system.

**FPGA platform characteristics:** The FPGA platform utilized in the following tests is based on Xilinx Virtex II Pro XC2VP30 chip. It's a multitude of feature among them the ability to host software processes such as MicroBlaze (MB) or also the possibility to exploit hardware processors already available such as the PowerPC processor. It contains up to 30000 logic blocks offering the possibility to incorporate in the design many custom IPs. There are a miscellaneous set of predefined IPs in the Xilinx library such as GPIOs, timers, memory blocks and many others. On the communication side, designers have the choice between different communication buses like On Chip Peripheral Bus (OPB) or the Processor Local Bus (PLB). It is also possible to utilize point to point links via Fast Simplex Links (FSL). It is a 32 bits width connection and the data time access consumes between 1-2 clock cycles. In the tests conducted later, FSL point to point links are used for performance reasons to connect the hardware module of the DCM emulator as a coprocessor to the MB processor.

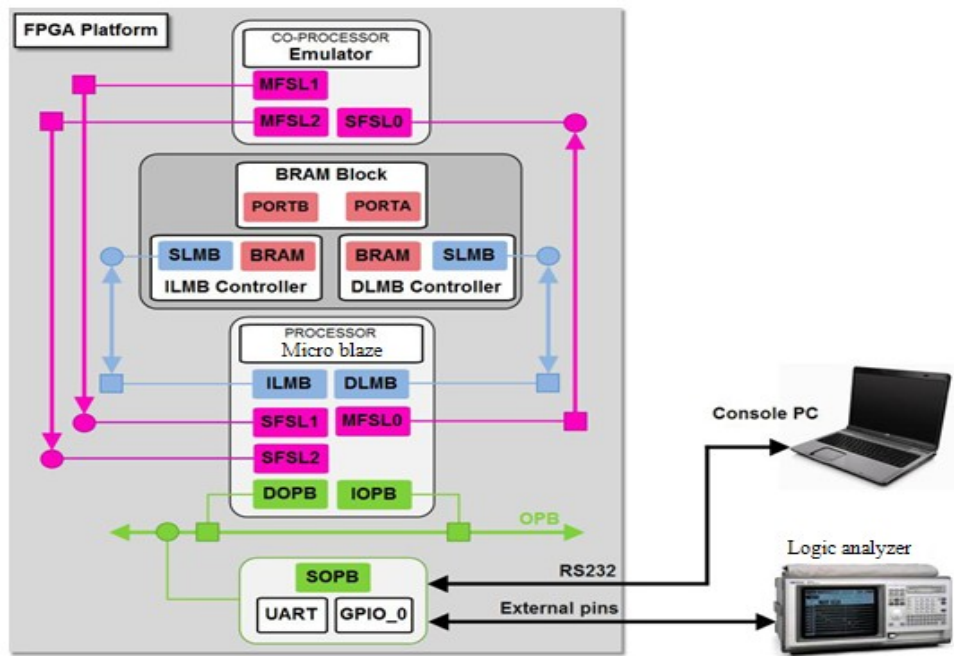

Fig. 10: System architecture (Design 1)

**Design 1: Open loop system test:**
**System architecture:** When the HLS methodology is used in a design flow, designers may use the resulting hardware modules without any modification. However, this doesn't exclude the possibility to refine the final circuit. In this case study, we added to the hardware module of the emulator some ports supported by GPIO external pins. In fact, they are linking internal debug signals which are automatically generated by CoDeveloper to the logic analyzer. These new connections were very helpful; they allow us the real time to watch data transfer, to elaborate timing measurement and eventually to detect emulator abnormal functioning.

After this port modification, the emulator hardware module is associated with the MB processor via FSL links and the complete system is implemented in the FPGA card. The hardware architecture is summarized in Fig. 10. On the software side, the drivers used for read/write operations from/to the emulator circuit consist of simple Put/Get FSL instructions as shown below:

- Putfsl (alpha, 0) : alpha_data is sent to the emulator via FSL_0.

- Getfsl(Im,1) : im_data and wm_data are getfsl(Wm,2) received by the processor via FSL_1 and FSL_2.

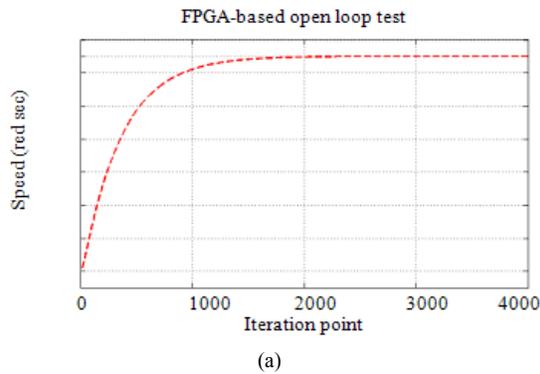

(a)

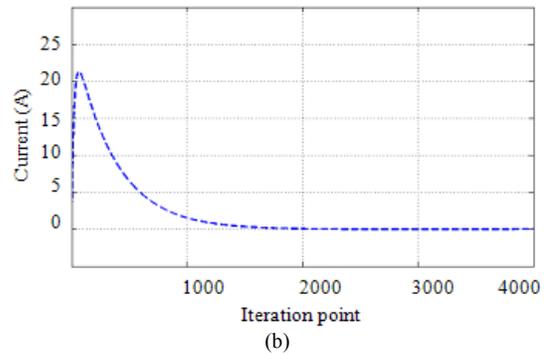

(b)

Fig. 11: FPGA-based Speed (a) and Current (b) responses of the DCM emulator (open loop)

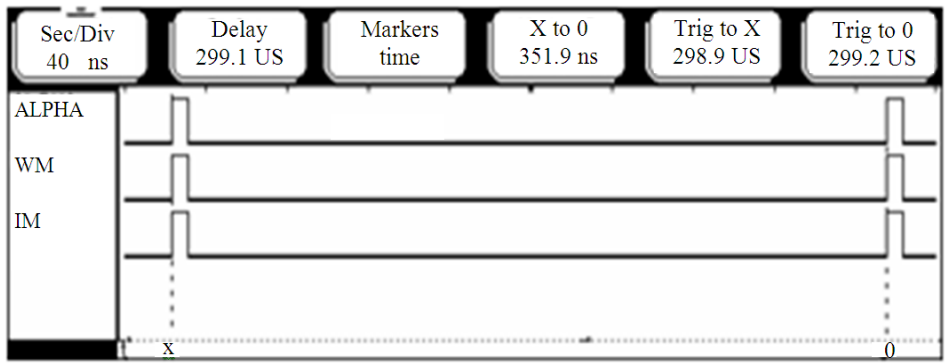

Fig. 12: FPGA-based emulator execution time

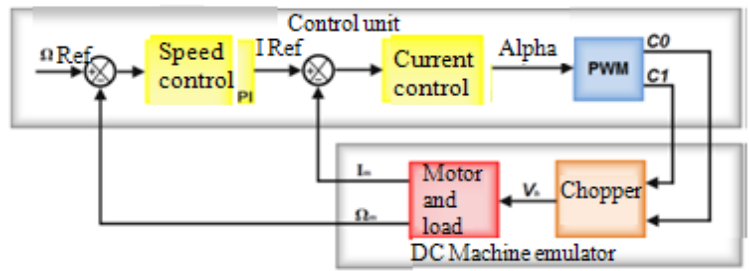

Fig. 13: DCM process diagram with PWM module

**FPGA-based DCM emulator execution time and computing results:** The communication between the FPGA board and the computer is assured by a serial connection via a UART module. It allows us to obtain data from the emulator and to analyze them. Im and Ωm data were, first saved in a local memory and just after the end of the saving operation, sent to the console PC and stored in an output file to be analyzed (Fig. 10). Internally, the alpha level used is 0.5-0.7 at zero and sent to the emulator via the FSL-0 link. We have recuperated a 4k values issued from an incremental iteration of the DCM emulator algorithm. This results in the evolution curves illustrated in Fig. 11.

An important characteristic of an emulator is the capability to reproduce the motor model. We can recognize the perfect coherence between curves in Fig. 8 and Fig. 11. The FPGA-based simulation matches closely the PC-based simulation (realized by Application Monitor). The execution time is measured with the logic analyzer from GPIO pins. Fig 12 shows a screenshot of the measure operation.

We can see in the figure above the execution time corresponds to which was measured in the simulation stage (page 7).

**Design 2: Hardware-in-the-Loop test System architecture:** In this design, we have evaluated a digital controller for DC motor. For more accuracy and to be closer to a real DC machine process, we have added a PWM module which controls the chopper by two opposite and logic signals (C0 and C1). Fig. 13 shows the system architecture after this modification.

As it is shown in Fig. 13, the inputs of the emulator have changed because it will no longer be connected directly to a data bus that furnishes the duty cycle alpha. In this design, the emulator will acquire C0 and C1 signals (1 bit width each one). In the algorithm side, Eq. 4 is replaced by this equivalent code developed inside the emulator for the chopper where the new input ports are considered:

```
if ((C0 == 1) and (C1 == 0)) Vh = Vin ;
else
{ if ((C0 == 0) and (C1== 1)) Vh = -Vin ;
else (Vh = 0; }
```

The substitution of Eq. 4 of the last code induces some ramifications in the hardware module of the emulator. HLS technique makes these modifications possible and quicker because it enables a faster switch between configurations without any dependency on a hardware design specialist. This point will be detailed later in discussion.

To complete the HiL test, a controller was added to the system design. The controlling algorithm is computed in a software manner. It is assured by two interruptions, respectively for current and speed controls and which are executed in the MicroBlaze processor. The hardware architecture for design 2 is globally similar to design 1 except for some modifications which are caused by the addition of the control unit. They are listed below:

- Suppression of FSL-0 data bus
- The addition of the PWM module as a slave OPB (alpha is sent to this module via this connection)
- Connection of C0 and C1 ports to the PWM through external pins
- The addition of 2 Timers to schedule the interruption functions of current and speed controllers
- Addition of an Interrupt Controller

**HiL test results:** The emulation system was executed in 1.5 seconds on the FPGA card with a speed reference $\Omega_{Ref}$ of 100 rad/s. The emulator rotation speed and current responses are shown in Fig. 14.

We can see in Fig. 14 that reference speed (100 rad/s) was correctly reached in stationary mode. For the current curve, there are many fluctuations that are not present in the speed curve. They are the result of these factors:

- The current is more sensitive to the addition of the PWM module in this design than the speed

For the saving operation, we opted for a (MB + emulator) architecture where the MB processor is charged of gathering data from the emulator. We scheduled an interruption which occurs each 15μ (fastest time with the MB processor) to save values in local memory. Due to the emulator"s execution time of 1μs, we were able to save one value from fifteen each saving operation. This is why the obtained results will only permit the validation of the shape of the current curve in Fig. 14. A more complex and enhanced solutions are conceivable in the future for better saving operation.

Note that execution time mentioned in the preceding point 2) is 1 μs although we noticed in the simulation (page 9) and on-chip measure (page 10) that this time is about 350 ns. In fact, the execution time of the emulator was slowed down to meet the objective of 1 μs fixed at the beginning of the study. Technically, it consists of the addition of a timer with a logic output signal with a period of 1 μs. This signal is connected to the enable ports of the emulator (xxx_en, see Fig. 7) to control its functioning. These results are exploited in the following paragraph to demonstrate the advantages behind the hardware implementation of the DCM emulator.

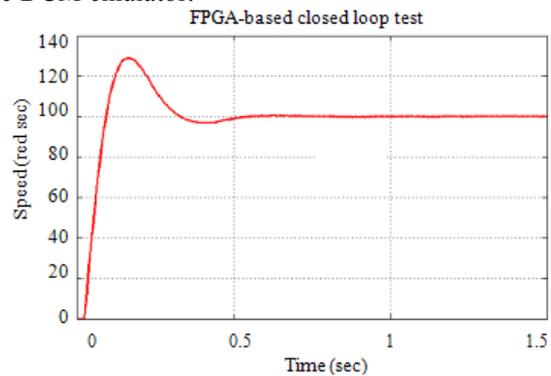
(a)

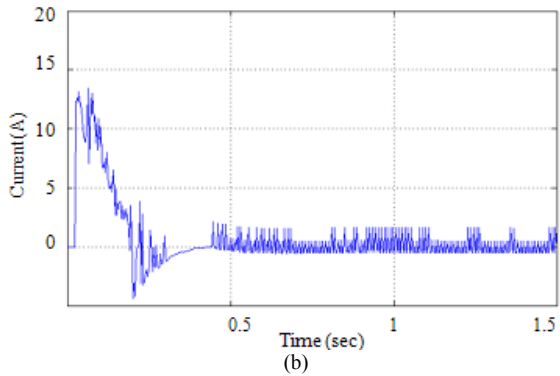

Fig. 14: FPGA-based speed (a) and current (b) responses of the DCM emulator (closed loop)

**Controller failures insertion and detection:** in this test, three controller failures (Fault1, 2, 3) were realized and inserted into the controller unit. They correspond to 15, 30 and 60 μs of PWM module inactivation time where C0 and C1 signals were held to zero logic value as shown in Fig. 15. This operation is equivalent to a typical controller fault that is the execution stopping of the control algorithm. These faults are common in the automotive domain where the lack of supply voltages occurs frequently and in short delays. This exercise was very useful because it permitted, at the same time, the validation of all the system components (Emulator, Controller, Interconnections, Timers, PWM module) in real time functioning. The resulting current curves for each fault were plotted and compared to the normal one in Fig. 16.

Results in Fig. 16 shows that each current curve has adopted its proper trajectory when the faults were cleared. This is because the faults were inserted into closed loop mode so the emulator and its controller were affected, but the most important resides in the fact that this difference in current responses demonstrates that these controller faults were detected by the DCM emulator.

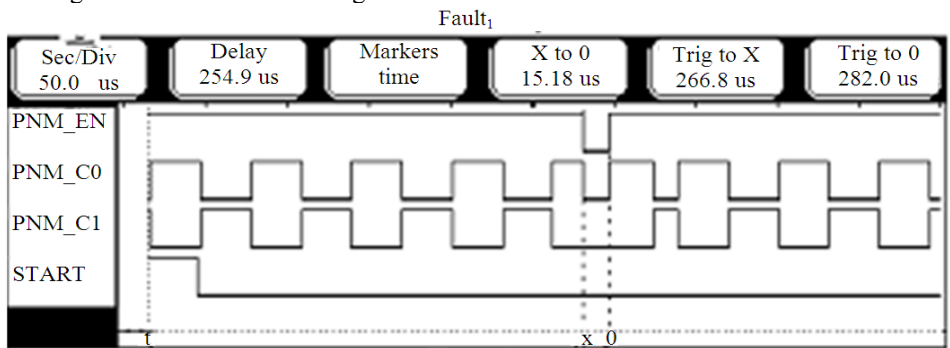

(a)

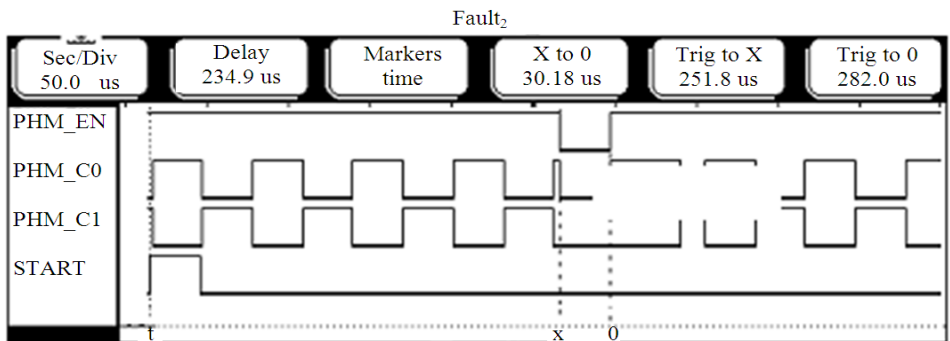

(b)

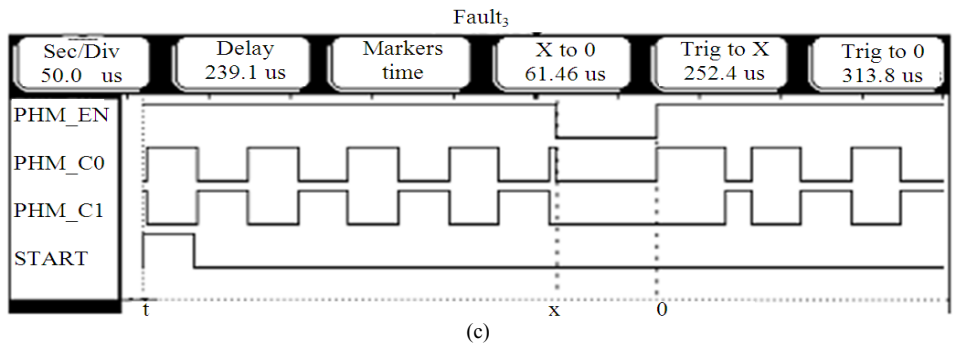

Fig. 15: Occurrence and clearance of fault 1 (a), 2 (b) and 3 (c)

This test was sufficient to explain that the detection of such rapid command faults (a few microseconds) is a direct result of the RT DCM emulator high computing speed.

**DISCUSSION**

In this paragraph, we will discuss two major points developed throughout this case study:

- Advantages and disadvantages behind the application of HLS to the DC control domain
- The results obtained from the hardware implementation of the DCM emulator

Generally, HLS technique accelerates the design flow by avoiding hardware conception bottlenecks in two phases of the hardware conception stage:

- At the beginning of the design: If the application is new, designers have to concept all the system and cannot make benefits from older experiences or using the previous available Intellectual Property (IP) cores
- At the end of the design: If the final circuit has to be modified it may induce the re-design of important parts of the system

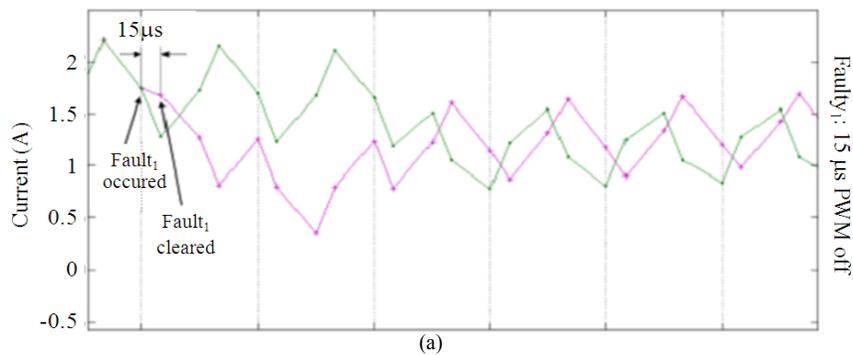

(a)

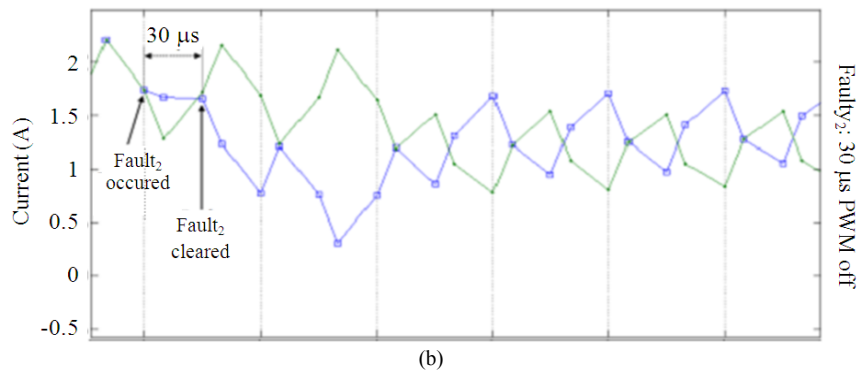

(b)

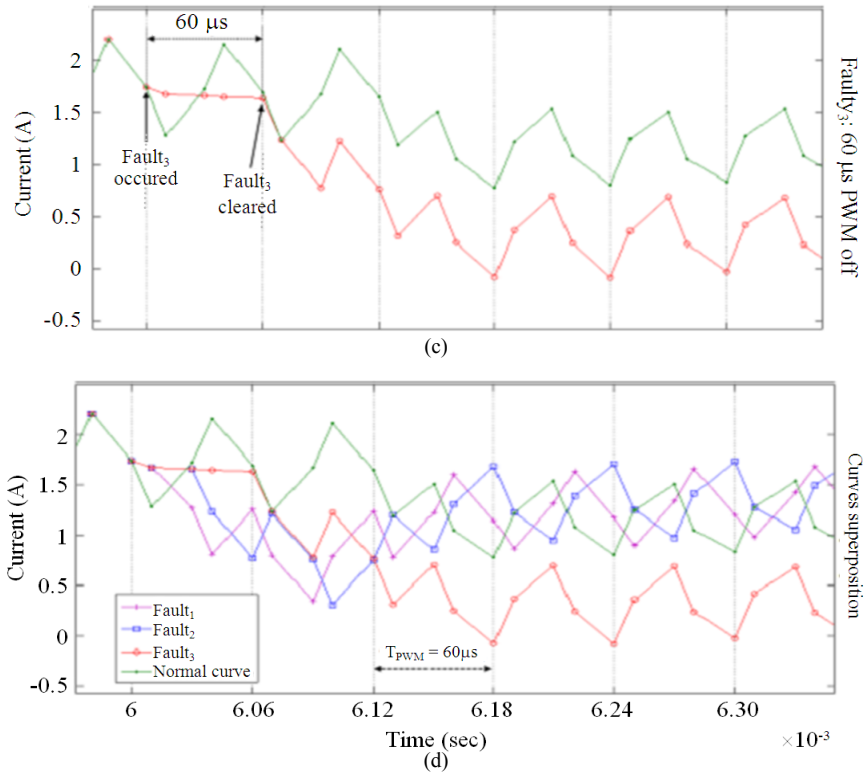

Fig. 16: Emulator currents responses after faults insertion -15 µs (a), 30 µs (b), 60 µs (c)

Table 5: Design time comparison

|  | First attempts (New design) | Modifications in the final design |
|---|---|---|
| Hand coded VHDL (Estimation) | 4 days | 2 days |
| HLS approach | 2 hours | 10 mn |
| Design acceleration | 48 × | 80 × |

For the design of the emulator, we took advantage from both of the phases cited above. Table 5 shows a comparison between HLS based design and hand coded VHDL for the emulator module used in this case study for the first design and after the modifications done in Design 2 (page 9).

The design acceleration obtained by HLS is important and promising. However, when we consult the hardware synthesis result in Table 6 we can clearly conclude that area consumption for the emulator core is excessive. Moreover, it may overload the targeted FPGA circuit in case of more complex models (e.g., AC motors). In this study, although the area consumption was not a critical constraint like the computing time, we propose some improvements which can help to avoid this issue. These optimizations will induce the design of a more refined hardware description which considers hardware constraints and keeps a competitive execution time:

- The optimization of the considered algorithm (Algorithm dependent)
- The conversion of the algorithm computing from floating representation to fixed point (Good)
- The use of tool-specific optimization commands such as CO PIPELINE (Average)

The computing time obtained from moving the emulator software algorithm to a hardware implementation is useful in the DC control domain. In fact, it offered a more accurate model for the validation and diagnosis of DC motor controllers by permitting tests that were not realizable with software emulators. To conclude this section, we expose a resume of our previous results in this domain (Table 7).

Table 6: Hardware synthesis report

| Hw Resources | Micro blaze | DC Motor emulator (Design 2) | PMW | Timers | Others | FPGA resources utilization |
|---|---|---|---|---|---|---|
| Slices | 1765 | 2366 | 198 | 552 | 373 | 38 % |
| Slice flip flops | 1886 | 3377 | 225 | 490 | 128 | 22 % |
| 4 inputs LUTs | 2720 | 2884 | 235 | 398 | 403 | 24 % |
| Mult 18×18 | 7 | 16 | - | - | - | 17 % |

Components

Table 7: Classification of the previous results for the DCM emulator

| Reference | System architecture | | DC Motor emulator Execution time | Acceleration versus 350 ns |
|---|---|---|---|---|
| | Configuration | Hw/Sw Partitioning | | |
| (Ben Othman *et al.*, 2008) | MPSoC (3 Micro Blaze processors) | 3 Interruptions (2 controllers + emulator) each one running in its proper processor | 138 μs | 394× |
| (Ben Salem *et al.*, 2008) | Single PowerPC + μC/OS-II RTOS | 3 tasks (2 controllers + emulator) | 22 μs | 63× |
| (Ben Salem *et al.*, 2010) | Single MicroBlaze + μC/OS-II RTOS + FPU unit + emulator) | 3 tasks (2 controllers | 900 ns | 2.5× |
| This work | Single MicroBlaze + Co-processor + PWM (hw) | 2 Interruptions (2 controllers) + hardware emulator | 350 ns | - |

## CONCLUSION

In this study, a hardware conception and implementation of a Real Time Direct Current Machine emulator was realized. Conception stage was conducted using HLS technique benefits to avoid VHDL manual work and to accelerate the design flow. As a result, a high accurate emulating system with a very low sampling rate that allows the high fidelity representation of a real DC Motor. Timing analysis, hardware simulation and on-chip verifications demonstrates that the use of this emulating system as a practical validation stage for recent DC digital controllers is possible.

As mentioned throughout this case study, control algorithms and electrical motor models become more and more complex, so the application of HLS technique in this domain can be limited in the future especially by the size of generated hardware IPs. Paradoxically, the area consumption may not be a limitation in the near future thanks to the next-generation of FPGA platforms which offers a multitude of advantages among them the high integration capacity and the reduced power consumption. As an example, we cite devices based on 40 nm and 28 nm technologies which are already commercialized (Chen *et al.*, 2010). Besides, a hardware implementation of all the DCM process, including the controller is feasible in the future.

Moreover, the application of HLS technique to a more complex electrical machine model like Asynchronous Machine is also conceivable. However, as it was shown in discussion paragraph, this technique should be used carefully with control algorithms and the compromise between accelerating the workflow time and the quality of the final design (Area consumption versus computing speed) has to be studied in advanced stages of design to avoid unsuitable results.

# REFERENCES

Ben Achballah, A., S. Ben Othman and S. Ben Saoud, 2010. High Level Synthesis of Real Time Embedded Emulating System for Motor Controller. IEEE International conference on Design and Technology of Integrated Systems in Nanoscale Era, Mar. 23-25, IEEE Xplore Press, Tunisia, pp: 1-6. DOI: 10.1109/DTIS.2010.5487549

Ben Othman, S., A.K. Ben Salem and S. Ben Saoud, 2008. MPSoC Design of RT Control Applications based on FPGA Soft Core Processors. IEEE International Conference on Electronics, Circuits and Systems, Aug. 31- Sept. 3, IEEE Xplore Press, Tunis, pp: 404-09. DOI: 10.1109/ICECS.2008.4674876

Ben Salem, A.K., S. Ben Othman and S. Ben Saoud, 2008. RTOS for SoC Embedded Control Applications. IEEE International conference on Design and Technology of Integrated Systems in Nanoscale Era, Mar. 25-27, IEEE Xplore Press, Tunis, pp: 1-6, DOI: 10.1109/DTIS.2008.4540273

Ben Salem, A.K., S. Ben Othman and S. Ben Saoud, 2010. Field Programmable Gate Array -Based System-on-Chip for Real-Time Power Process Control. Am. J. Applied Sci., 127-39, DOI: 10.3844/ajassp.2010.127.139

Ben Saoud, S., B. Dagues, H. Schneider, M. Metz and J.C. Hapiot, 1996. Real time emulator of static converters/electrical machine application to the test of control unit. Proceedings of the IEEE International Symposium on Industrial Electronics, Jun 17-20, IEEE Xplore Press, Toulouse, pp: 856-861. DOI: 10.1109/ISIE.1996.551055

Braham, A., H. Schneider and M. Metz, 1997. Recent developments in real-time analogue simulation of high power electro-technical systems. Proceedings of the 23rd International Conference on Industrial Electronics, Control and Instrumentation, Nov 9-14, IEEE Xplore Press, Toulouse, pp: 744-748. DOI: 10.1109/IECON.1997.671918

Dufour, Christian, Belanger, Jean and Lapointe, Vincent 2008. FPGA-based Ultra-Low Latency HIL Fault Testing of a Permanent Magnet Motor Drive using RT-LAB-XSG. Proceedings of the Joint International Conference on Power System Technology and IEEE Power India Conference, Oct. 12-15, IEEE Xplore Press, pp: 1-7, DOI: 10.1109/ICPST.2008.4745355

Dufour, C., Belanger, J., S. Abourida and V. Lapointe, 2007. FPGA-based real-time simulation of finite-element analysis permanent magnet synchronous machine drives. IEEE Power Electronics Specialists Conference, June 17-21, IEEE Xplore Press, Montreal, pp: 909-915. DOI: 10.1109/PESC.2007.4342109

Gonzalez, I., E. El-Araby, P. Saha, T. El-Ghazawi and H. Simmler, *et al.,*, 2008. Classification of application development for fpga-based system. IEEE National Aerospace and Electronics Conference, July 16-18, IEEE Xplore Press, Washington, pp: 203-208. DOI: 10.1109/NAECON.2008.4806547

Gupta, S., N. Dutt, R. Gupta, A. Nicolau, 2004. Loop shifting and compaction for the high-level synthesis of designs with complex control flow', Proceedings of the conference on Design, Automation and Test in Europe, Feb. 16-20, ACM, Washington, DC, USA, pp: 114-119. http://dl.acm.org/citation.cfm?id=969079

Idkhajine, L., A. Prata, E. Monmasson, K. Bouallaga and M.W. Naouar, 2008. 'System on Chip Controller for Electrical Actuator. IEEE International Symposium on Industrial Electronics, June 30-July 2, IEEE Xplore Press, Cergy, 2481-2486. DOI: 10.1109/ISIE.2008.4677002

Martin, G. and G. Smith, 2009. High-level synthesis: Past, present and future. IEEE Design and Test of Computers, (IDTC' 09), IEEE Xplore Press, USA, pp: 18-25. DOI: 10.1109/MDT.2009.83

Monmasson, E. and M.N. Cirstea, 2007. FPGA design methodology for industrial control systems -- a review. IEEE Trans. Industrial Electronics, 54: 1824-1842. DOI: 10.1109/TIE.2007.898281

Naouar, M.W., E. Monmasson, A.A. Naassani, I. Slama-Belkhodja and N. Patin, 2007. FPGA-Based Current Controllers for AC Machine Drives—A Review. IEEE Trans. Industrial Electronics, 54: 1907-1925. DOI: 10.1109/TIE.2007.898302

Ouhrouche, M., 2009. Transient analysis of a grid connected wind driven induction generator using a real-time simulation platform. Renewable Energy, 34: 801-806. DOI: 10.1016/j.renene.2008.04.028

Paiz, C., C. Pohl and M. Porrmann, 2008. Hardware-in-the-Loop Simulations for FPGA-based Digital Control Design. Informatics Control Automation and Robotics 15: 355-3720. DOI: 10.1007/978-3-540-79142-3_27

Pellerin, D. and S. Thibault, 2005. Practical FPGA Programming in C. 1st Ed. Prentice Hall, ISBN: 0131543180, pp: 464.

Rodriguez-Andina, J.J., M.J. Moure and M.D. Valdes, 2007. Features, Design Tools and Application Domains of FPGAs. IEEE Trans. Industrial Electronics, 54: 1810-1823. DOI: 10.1109/TIE.2007.898279

Salewski, F. and S. Kowalewski, 2008. Hardware/Software Design Considerations for Automotive Embedded Systems. IEEE Trans. Industrial Informatics, 4: 156-63. DOI: 10.1109/TII.2008.2002919.